\documentclass{emulateapj}

\slugcomment{Accepted for publication in ApJL}

\shorttitle{Cluster candidates at $z \sim 1.8$}
\shortauthors{Chiaberge, M., et al.}

\begin{document}


\title{Three candidate clusters of galaxies at redshift $\sim1.8$: the
``missing link'' between protoclusters and local clusters?}


\author{Marco Chiaberge\altaffilmark{1,2}, A. Capetti\altaffilmark{3}, F. Duccio Macchetto\altaffilmark{1}, P. Rosati\altaffilmark{4}, P. Tozzi\altaffilmark{5}, and G.~R. Tremblay\altaffilmark{6} }

\altaffiltext{1}{Space Telescope Science Institute, 3700 San Martin Drive, Baltimore, MD 21218}
\altaffiltext{2}{INAF - IRA, Via P. Gobetti 101, Bologna, I-40129}
\altaffiltext{3}{INAF - Osservatorio Astronomico di Torino, Via Osservatorio 20, I-10025 Pino Torinese, Italy}

\altaffiltext{4}{European Southern Observatory, Karl-Schwarzschild-Strasse 2, D-85748, Garching bei M\"unchen}
\altaffiltext{5}{INAF - Osservatorio Astronomico di Trieste,  Via Tiepolo 11, I-34143 Trieste, Italy}
\altaffiltext{6}{Rochester Institute of Technology, One Lomb Memorial Drive, Rochester, NY 14623 USA}



\begin{abstract}

We  present three  candidate clusters  of galaxies  at  redshifts most
likely  between 1.7  and  2.0, which  corresponds  to a  fundamentally
unexplored epoch of clusters  evolution.  The candidates were found by
studying  the   environment  around  our  newly   selected  sample  of
``beacons''  low-luminosity  (FR~I)   radio  galaxies  in  the  COSMOS
field. In  this way we intend  to use the fact  that FRI at  low z are
almost invariably  located in clusters  of galaxies.  We use  the most
accurate  photometric  redshifts available  to  date,  derived by  the
COSMOS collaboration  using photometry  with a set  of 30  filters, to
look   for   three-dimensional   space   over-densities   around   our
objects.  Three out  of the  five FR  Is in  our sample  which possess
reliable photometric redshifts  between $z_{phot}=1.7$ and 2.0 display
overdensities  that  together  are  statistically significant  at  the
4$\sigma$ level, compared to field counts, arguing for the presence of
rich  clusters of  galaxies  in their  Mpc  environment.  These  first
results  show that  the new  method for  finding high-$z$  clusters we
recently proposed,  which makes use  of low power FR~I  radio galaxies
instead  of  the  more  powerful  FR~II  sources  often  used  in  the
literature to date, is returning very promising candidates.

\end{abstract}



\section{Introduction}

The  search  for  clusters of  galaxies  at  $z>1$  has proven  to  be
particularly difficult, mainly because of the reduced contrast between
cluster  members  and  field  galaxies. High-$z$  clusters  have  been
searched for  using a variety  of techniques, including  the so-called
``red  sequence'' technique  \citep{gladders00}, and  the use  of both
X-ray surveys \citep[][and references therein]{rosati02}, and infrared
surveys   \citep[e.g.][]{elston06}.    Recently,  \citet{eisenhardt08}
presented a sample  of clusters of galaxies from  the Spitzer Infrared
Array Camera Shallow Survey.  Of their 335 cluster candidates in their
sample,  12  have  confirmed  spectroscopic  redshift  $1  \lesssim  z
\lesssim  1.4$, and  a few  (less than  10) have  photometric redshift
(obtained using photometry in 4  bands, B$_W$, R, I, and 3.6$\mu$m) in
the  range  $1.5  \lesssim  z_{phot}  \lesssim 1.7$.   None  of  their
candidates  have either photometric  or spectroscopic  redshift higher
than 1.7.   A similar  effort is being  pursued by the  SpARCS project
\citep[e.g.][]{wilson09},  which found  no candidates  at $1.4  <  z <
2.2$.

Radio  galaxies  \citep[e.g.][]{hall01,best03,venemans07} and  quasars
\citep[e.g.][]{haas09} have also been  extensively used to find high-z
clusters  \citep[see   also][for  a  review]{miley08},   as  they  are
associated with massive elliptical  galaxies and clusters in the local
universe.  However,  it is known  that only low power  radio galaxies,
most of  which possess  an ``edge-darkened'' radio  morphology, namely
FR~I \citep{fanaroffriley}, are  almost invariably located in clusters
of  galaxies \citep[e.g.][and  references therein]{zirbel96,owenrev96}
and   are  also   associated  with   the  brightest   cluster  members
\citep[e.g.][]{best07}. For $z>0.5$ an increasing fraction of the more
powerful   ``edge-brightened''   FR~IIs  are   also   found  in   rich
environments\citep[e.g.][]{prestagepeacock88,hilllilly91,best00},   but
the fraction  of FR~II in  rich clusters is  still lower than  that of
FR~Is.  Furthermore,  the host galaxies  of FR~Is are more  similar to
normal   ``inactive''  ellipticals  than   those  of   FR~IIs.   Giant
Ly$\alpha$ emitters, possibly  representing protocluster regions, have
been   discovered   around    powerful   radio   galaxies   at   $z>2$
\citep[e.g.][]{steidel00,pentericci01,zirm05},  but  the  relationship
with today's clusters  is still a matter of  debate, mainly because of
the small sample  of clusters known between $z\sim 1$  and $2$, and in
particular at $1.5 < z <2$.

Here we take  advantage of the specific properties  of the environment
of FR~Is to  set the stage for filling that  redshift gap.  We utilize
our newly  selected sample of  high-$z$ low luminosity  radio galaxies
\citep{paphighzfr1} (hereinafter Paper~I)  as ``beacons'' for clusters
in the unexplored range of redshift.  In this Letter, we present first
results that  returned three  very promising cluster  candidates, most
likely located at $z\sim 1.7 - 2.0$.

We use  the following  cosmological parameters
\citep[$H_0   =71$   Km   s$^{-1}$   Mpc$^{-1}$,   $\Omega_M=   0.27$,
$\Omega_{vac}= 0.73$,][]{hinshaw09}.

\begin{figure*}
\epsscale{1.2}
\plotone{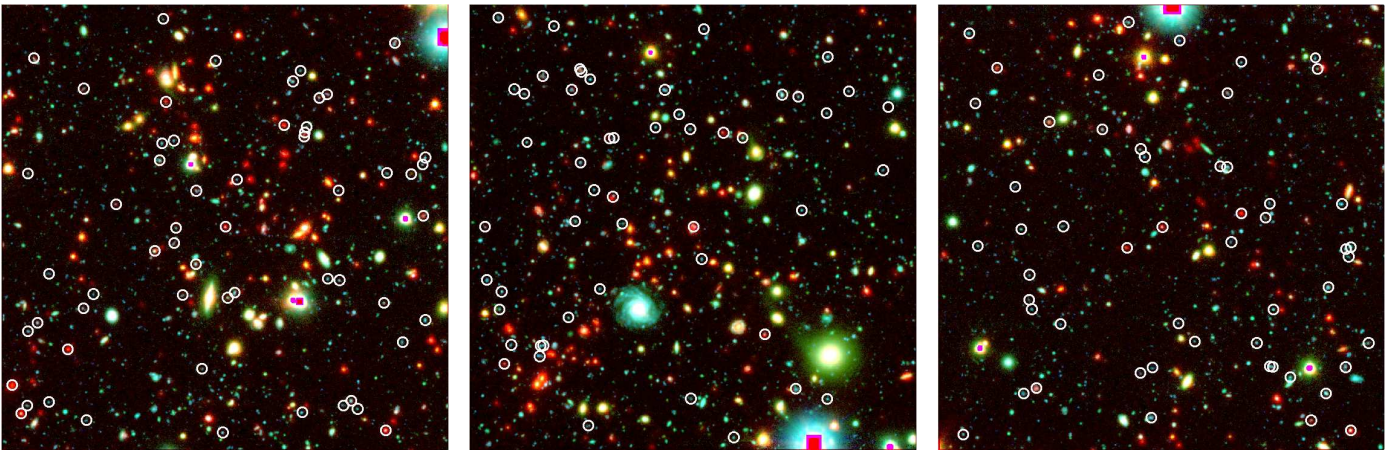}
\caption{RGB ``color''  images of the three  candidate clusters around
COSMOS-FR~I~03 (left),  05 (center), and 226 (right).   The images are
obtained  using {\it  Spitzer} 3.6$\mu$m,  {\it Subaru}  r and  B band
images for the R, G and B channels respectively.   White circles
indicate objects  with $1.6< z_{phot}  < 2.3$.  The projected  size of
the fields are $180\arcsec  \times 180\arcsec$ ($\sim
1.5 \times 1.5$ Mpc$^2$ at $z\sim 1.8$), North is up.}
\label{rgb}
\end{figure*}

\section{The sample of FR~I ``beacons''} 

Details of  the selection  of the high-$z$  ($1<z<2$) FR~I  sample are
given in  Paper~I.  Here we briefly  summarize the relevant
steps.  

We  proceed in four  steps, under  the basic  assumptions that  1) the
FR~I/FR~II break  in radio  power per unit  frequency \citep[usually  set at
$L_{1.4 GHz} \sim  4 \times 10^{32}$ erg s$^{-1}$  Hz$^{-1}$][]{fanaroffriley} does not
change with redshift, and that 2) the magnitude and color of the hosts
of FR~Is at $1 < z < 2$ are similar to those of FR~IIs within the same
redshift   bin,   as   in   the   case   of   local   radio   galaxies
\citep[e.g.][]{donzelli07}.  Note that the photometric redshift is not
a selection constraint.

1) We select FIRST \citep{becker95}  radio sources in the COSMOS field
\citep{scoville07}, whose  1.4 GHz fluxes  correspond to the  range of
fluxes expected for FR~Is at $1<z<2$ ($1<F_{1.4} < 13$ mJy).

2) Sources with FR~II radio morphology are excluded.

3) Bright  (m$_I >  22$) galaxies  are  rejected since  they are  most
likely  low$-z$  galaxies  with  faint radio  emission  (e.g.   nearby
starbursts).

4) U-band dropouts  are rejected as they are  likely to be at  $z > 2.5$.  

We are  left with  a sample of  37 FR~I  candidates, all of  which are
identified in the COSMOS  catalog \citep{capak07}.

\section{Using FR~I to find high-$z$ clusters}
\label{find}

To  confirm  that  the  FR~I sources  reside  in  a  cluster
environment at  $z>1$, spectroscopic  redshifts of the  target itself,
and eventually of a significant  number of galaxies in the surrounding
region of the  sky would be ideal.  
Unfortunately, the $z$-COSMOS  catalog \citep{zcosmos} does not include
our sources.

In absence  of spectroscopic data,  we take advantage of  the recently
published  catalog of  photometric redshifts  obtained using  30 bands
photometry by  \citet{ilbert09} to identify the  best candidates.  The
catalog,  which includes  objects with  I  $ <  25$ in  the deep  {\it
Subaru}  area of  the  COSMOS field  \citep{taniguchi07} represents  a
significant    improvement    with   respect    to    the   list    of
\citet{mobasher07}, providing a factor  of $\sim 3-5$ higher accuracy.
The authors  estimate the redshift  accuracy to be  $\sigma_{\Delta z}
\sim 0.04$ for  galaxies of $i \sim 25$  at $z<1.25$.  For $1.5<z<2.5$
and $i \sim  25$ (similar to our faintest FR~I  hosts) the accuracy is
$\sigma_{\Delta z} \sim 0.19$, still  well suitable for the purpose of
this Letter.  Note that the spectral energy distribution of FR~Is host
galaxies  is  not   dominated  by  strong  emission  lines\footnote{We
estimate a  maximum contamination from the  strongest optical emission
line  (H$\alpha$) of  $\sim  1\%$.  The  narrowband  filters might  be
affected  by UV emission  lines (CIV~1549  and Mg~II~2800),  but their
contribution  is  not  expected  to  exceed  $10\%$  \citep{sara09}.},
contrary to  those of  all other AGNs.   The galaxy templates  used by
\citet{ilbert09} are therefore adequate for our analysis.

We search the \citet{ilbert09} catalog for the 37 sources presented in
Paper~I
and we find all of them, but one.  Four have $1.4 < z_{phot} <
1.7$  and 5 have  $1.7 \leq  z_{phot} <  2.0$.  Here  we focus  on the
latter  objects, since  they are  among the  highest  redshift cluster
candidates known to date.  

\begin{table}
\begin{center}
\caption{Photometric redshifts\label{table}}
\begin{tabular}{l c c c c}
\tableline\tableline
Source & $z_{phot}^a$ & 99\% range$^a$ & 68\% range$^a$ & $z_{phot}^b$\\
\tableline
COSMOS-FR~I~03     &  1.96  & 1.55-2.32 & 1.88-2.15 & 1.59\\
COSMOS-FR~I~05     &  1.84  & 1.72-2.06 & 1.82-1.87 & 2.08\\
COSMOS-FR~I~22\tablenotemark{a}   &  1.79  & 1.69-2.08 & 1.74-1.82 & 1.50 \\
COSMOS-FR~I~226    &  1.76  & 1.62-2.37 & 1.71-2.03 & 2.04 \\
COSMOS-FR~I~228\tablenotemark{a}  &  1.88  & 1.30-2.63 & 1.81-2.05 & 1.45 \\
\tableline
\end{tabular}
\tablenotetext{a}{Not    shown   in    this    Letter   (see    text)}
\tablecomments{(1) =  source name, (2) = best  fit photo-$z$ estimate,
(3) and (4)  99\% and 68\%  confidence level range of  photo-$z$, $a=$
\citet[from][]{ilbert09}.   Column   (5)   =  photo-$z$,   $b=$   from
\citet{mobasher07}}.
\end{center}
\end{table}

The five  objects with $1.7 \leq  z_{phot} < 2.0$ are  objects 03, 05,
22, 226 and 228 in the  catalog of Paper~I.  We refer to these objects
as  ``COSMOS-FR~I~nn''.  The \citet{ilbert09}  catalog gives  a ``best
fit'' $z_{phot}$ and  a range of probable redshifts,  at 68\% and 99\%
confidence level (see Table~\ref{table}, where we also compare these
photo-$z$   with  the   previous  estimates   by  \citet{mobasher07}).

Two out  of the five  objects (COSMOS-FR~I~03 and  COSMOS-FR~I~05) are
detected in the X-ray band (as point sources)
both  in  the {\it  XMM}-COSMOS  \citep{hasinger07}  and  in the  {\it
Chandra}-COSMOS data \citep{elvis09}.

In  this Letter,  we focus  on three  objects,  namely COSMOS-FR~I~03,
COSMOS-FR~I~05,   and  COSMOS-FR~I~226,   which   display  significant
overdensities  of galaxies  in the Mpc-scale environment.

In Fig.~\ref{rgb}  we show RGB images  of a region  of $180\times 180$
square arcsec,  corresponding to  $\sim 1.5 \times  1.5$ Mpc$^2$  at a
redshift  $z \sim 1.8$,  centered on  each of  the three  radio galaxy
hosts.   
The images are obtained using  {\it Spitzer} 3.6$\mu$m, {\it Subaru} r
and B band images for the R, G and B channels respectively.  The white
circles   indicate   objects  with   $1.6<   z_{phot}   <  2.3$   (see
Sect.~\ref{clusters}). The  ``beacon'' FR~I radio galaxies  are at the
center of  each of  the three  fields.  Note that  the host  galaxy of
COSMOS-FR~I~03, which  in Paper~I  we reported as  ``non-detected'' in
the HST-COSMOS I-band images  \citep{koekemoer07}, is in fact detected
in the deeper {\it Subaru}-COSMOS i-band data.

\begin{figure*}
\epsscale{1.2}
\plotone{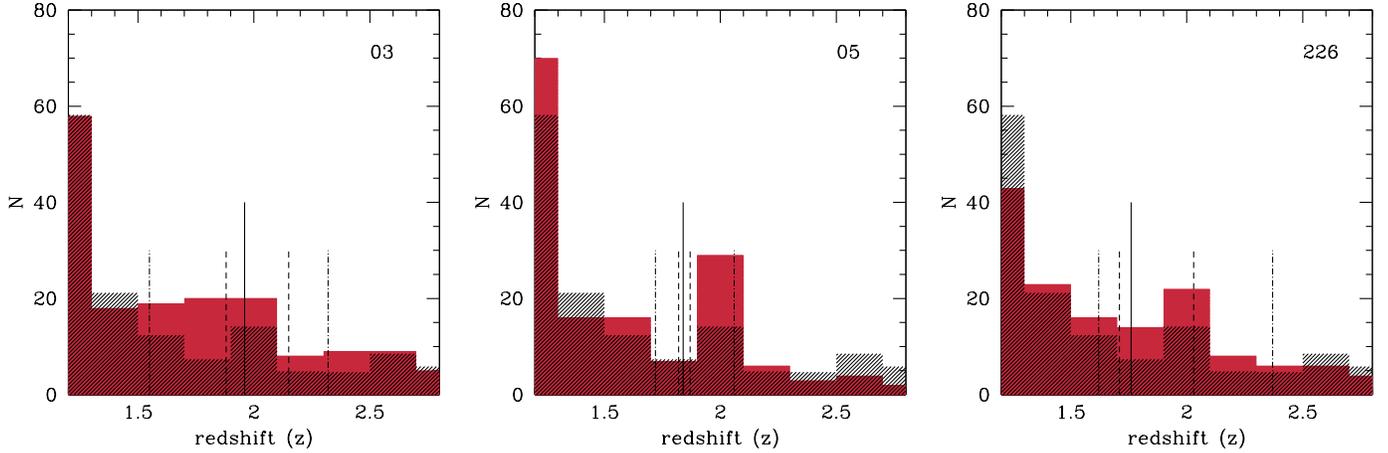}
\caption{Photo-$z$ distributions  of the $\sim$  Mpc scale environment
of   the   three   candidates   (red   histograms;
COSMOS-FR~I~03, 05 and 226 are shown in the left,
center  and right  panels, respectively)  as compared  to  the average
distribution  of the  control fields  (black shaded  histogram). 
Solid lines indicate the estimated  photo-$z$ of each of
FR~I ``beacon''.  Dashed and dot-dashed  lines show the  68\% and 99\%
range of confidence level, respectively (see Sect.~\ref{find}).}
\label{isto}
\end{figure*}

\vspace{10 mm}

\section{The cluster candidates}
\label{clusters}

In order to find more support  for the presence of cluster of galaxies
around   our  ``beacon''   radio   galaxies,  we   again  search   the
\citet{ilbert09}  catalog for  sources  that lie  within  a radius  of
90$\arcsec$ from  the FR~I host.  We  restrict the search to  a bin of
photometric  redshift $1.6  < z_{phot}  <  2.3$.  The  choice of  that
particular  redshift range is  motivated by  the uncertainties  on the
estimate of $z_{phot}$ outlined above.   We find 58, 51 and 53 objects
in  the  selected  3-dimensional   regions  around  the  three  FR~Is,
respectively.   

In Fig.~\ref{rgb} we mark with  white circles the galaxies with $1.6 <
z_{phot} < 2.3$.  The K-band magnitude of the marked objects is in the
range  K$_s \sim  20.5-25$ \citep{ilbert09}.   Note that  the expected
magnitude of a passively evolving L$^{\star}$ galaxy at $z\sim 1.8$ is
K$_s  \sim  21$  \citep[e.g.][]{strazzullo06}. The  ``central''  radio
galaxy host (always  located at the center of the  fields shown in the
figures)  has a  distinct ``red''  color i-K$_s$  $\gtrsim 3$  and the
K$_s$ magnitude is in the range 20.5-21.9 \citep{ilbert09}. In Paper~I
we  showed that  our  FR~I hosts  lie  on the  K-z  relation for  more
powerful radio galaxy hosts with evolved stellar populations.

Note that  a large  number of  ``red'' objects are  not marked  in the
figure.   A significant number  of those  objects (e.g.   the galaxies
$\sim 30\arcsec$  NW of COSMOS-FR~I~03)  are indeed only  bright in
the infrared, and do not make the optical selection cut to be included
in the \citet{ilbert09}  catalog.  These red objects are  as bright as
or slightly fainter than the radio galaxy host in K-band.  Their color
is i-K$_s$  $\gtrsim 3$, consistent with  evolved stellar populations,
and their  magnitude is in  the range K$_s$  $\sim$ $20.5-22$.   A few
other  galaxies, which  appear slightly  less ``red''  than  the radio
galaxy host (e.g.  the group in  the South-East corner of the field of
COSMOS-FR~I~05, are simply foreground ($z_{phot} \sim 0.8-1$) galaxies
with an old stellar population.

We also perform  the same search on 6  randomly selected fields inside
the COSMOS survey area.  For 3-dimensional regions of the same size as
the ones  described above, we find 28  to 35 objects in  the 6 control
fields, with a mean value of $32\pm 2$ and a median value of 32.  This
implies an over-density factor of  $\sim 1.7$ for the Mpc environments
of the  three selected FR~Is,  statistically significant at  the $\sim
4\sigma$  level,   when  the   three  candidate  cluster   fields  are
combined. We also  checked that doubling the number  of control fields
does not alter the results.

In  Fig.~\ref{isto} we  show histograms  representing  the photometric
redshift  distribution  of the  sources  within  a 90$\arcsec$  radius
around our radio galaxies  (red histograms), compared with the average
distribution of an identical area in the control fields. Clearly, this
assumes that the ``best'' values  of $z_{phot}$ are accurate to better
than  the width  of redshift  bins, which  therefore is  chosen  to be
sufficiently large ($\Delta z=0.2$,  similar to the photo-$z$ accuracy
for faint objects at $1.5 <z <2.5$).

The  figure shows  the overdensities  of the  cluster  candidates with
respect to the  control fields, arguing for the  presence of clusters.
The space  density enhancements appear  to be centered  between $z\sim
1.8-2.0$, thus very close to the FR~Is' best fit $z_{phot}$ (indicated
by the  solid line in each  plot).  For the fields  of COSMOS~FRI~ 03
and COSMOS~FRI~226 the distribution of the ``extra'' sources is spread
across a larger  redshift range ($\sim 1.6-2$).  This  is not entirely
surprising, because of the uncertainty of the photo-$z$ estimates.

We  checked that  the overdensities  at  $z_{phot} \sim  1.8$ are  not
``artificially''  generated by foreground  clusters coupled  with large
photo-$z$ errors at  faint magnitudes. In fact, they  are unrelated to
the presence of any concentration  of lower-$z$ objects.  On the other
hand, in  a few control  fields we observe overdensities  at $z_{phot}
\lesssim 1$,  which do not  artificially produce any  overdensities at
higher redshifts.

Finally, we note that the two  other sources in the range of photo-$z$
of interest  (namely COSMOS-FR~I~22  and COSMOS-FR~I~228) do  not show
significant overdensities with respect to the control fields, at least
in the range  $1.6 < z_{phot} < 2.3$. 
One  possibility   is  that  the   photo-$z$  for  these   objects  is
significantly offset.

\section{Conclusions}

We  identified  three cluster  candidates  with  $z_{phot} \sim  1.8$,
according  to the  photometric redshift  catalog  of \citet{ilbert09}.
The  possible existence of  clusters is  supported by  the significant
over-density  measured  on the  Mpc  scale  environment  of the  three
selected radio galaxies.  These first  results show that the method of
using  low luminosity  radiogalaxies  as ``beacons''  for clusters  is
returning  promising  candidates.  A  systematic  study  of the  whole
sample  of 37  radio galaxies  is under  way.  That  will allow  us to
determine the typical environment  of these high-z sources, and firmly
establish whether they are more  likely to be associated with clusters
than FR~IIs in the same range of redshifts.

We have submitted observational proposals to spectroscopically measure
the redshift  of the candidates.  If confirmed,  these would represent
extremely  important  structures that  would  contribute  to fill  the
existing gap between $z\sim 1.5$ and $z\sim 2$ in cluster studies.

A  study of the  fields with  high accuracy  photometry and  deep high
angular resolution  images (with the HST) is  mandatory to investigate
the spatial  distribution and the morphology of  the cluster galaxies,
to derive the photometric properties of a significant number of member
candidates,  and  assess  the   existence  (or  the  absence)  of  the
color-magnitude relation at these high redshifts.

In  summary, follow-up  observations of  these fields  from  the radio
through the X-rays, will allow us to tackle some of the most important
questions on the  origin of clusters of galaxies  and cluster members,
to investigate the link between protoclusters at $z>2$ and well formed
clusters  in the  nearby  universe and  to  study the  effects of  the
evolution of the intracluster gas on the radio structures.

\acknowledgments

We acknowledge the effort of the entire COSMOS team.  We thank A. Zirm
for insightful  comments on the  manuscript.  We also  acknowledge the
anonymous referee  for valuable suggestions that  greatly improved the
paper.



\end{document}